\documentclass[aps,prb,twocolumn,floats,showpacs,amsmath,amssymb,floatfix,%
superscriptaddress]{revtex4}
\usepackage[dvips]{color}
\usepackage{graphicx}
\usepackage{epsfig}
\usepackage{dcolumn}
\usepackage{bm}
\usepackage{amsmath}
\usepackage{amssymb}
\begin{document}
\title{Decoherence and dephasing in Kondo tunneling through double
quantum dots}
\author{M.N. Kiselev}
\affiliation{Institute f\"ur Theoretische Physik, Universit\"at
W\"urzburg, 97074 W\"urzburg, Germany}
\author{K. Kikoin}
\affiliation{Physics Department, Ben-Gurion University of the
Negev, Beer-Sheva 84105, Israel}
\author{Y. Avishai}
\affiliation{Physics Department, Ben-Gurion University of the
Negev, Beer-Sheva 84105, Israel} \affiliation{Department of
Applied Physics, University of Tokyo, Hongo Bunkyo-ku, Tokyo 113
Japan}
\author{J. Richert}
\affiliation{Laboratoire de Physique Th\'eorique, Universit\'e
Louis Pasteur and CNRS, 67084 Strasbourg Cedex, France}
\date{\today}
\begin{abstract}
We describe the mechanism of charge-spin transformation in a
double quantum dot (DQD) with even occupation, where a time
dependent gate voltage $v_g(t)$ is applied to one of its two valleys,
whereas the other one is coupled to the source and drain electrodes.
The Kondo tunneling regime under  strong Coulomb blockade may be
realized when the spin spectrum of the DQD is formed by the ground
state spin triplet and two singlet excitations. Charge
fluctuations induced by $v_g(t)$ result in transitions within the
spin multiplet characterized by the $SO(5)$ dynamical symmetry group. In
a weakly non-adiabatic regime the decoherence, dephasing and
relaxation processes affect Kondo tunneling. Each of these
processes is caused by a special type of dynamical gauge
fluctuations, so that one may discriminate between the decoherence
in the ground state of a DQD and dephasing at finite temperatures.
\end{abstract}
\maketitle
\section{Introduction}
Manipulating spins in quantum dots in order to achieve new ways of
processing information has been a subject of intense research in
the last decade. \cite{qucom} One of the basic building blocks for
quantum computing is a double quantum dot (DQD),\cite{dqd} where
the concepts of entanglement and phase accumulation may be
effectively modelled and  studied. \cite{entangl} It turns out
that the key problem, which hampers the progress in implementing
the quantum information storage and processing is the
impossibility of perfect isolation of a complex quantum system
from the environment, which results in the loss of quantum
coherence.\cite{deco} The phase coherence of tunnel transport
through a quantum dot can be effectively controlled in an
Aharonov-Bohm geometry \cite{abohm} and/or in the Kondo
regime.\cite{kondr} To understand the generic features of  the
decoherence phenomenon, various model situations are considered,
where dephasing and relaxation processes are controlled by
specific mechanisms of interaction with the external environment.
Among these processes the interaction with a phonon bath
\cite{gurv} and an electron liquid \cite{KNG,Paas} should be
regarded as typical examples.

The aim of this paper is to study a specific mechanism of
decoherence and dephasing, which stems from the violation of {\it
dynamical symmetry} in a DQD. This dynamical symmetry is an
intrinsic property of any quantum dot with even occupation, where
the low-energy spectrum of spin excitations is formed by singlet
(S) and triplet (T) states. \cite{KA01,Nova} Since the interaction
with a reservoir (electrons in metallic leads) violates the spin
conservation law, S/T transitions accompany the cotunneling
through DQD and thus contribute to the tunneling transparency and,
in particular, to the Kondo-type zero-bias anomaly in tunnel
conductance.

The physical system in which we
 will study decoherence and dephasing effects is a DQD occupied
by two electrons in a T-shape geometry (TDQD), where only one of
its two wells is in tunnel contact with the leads  (see Fig.
\ref{f.00}). As it is established, \cite{KA01,Nova} the electron states
in this kind of quantum dot may be
 tuned in such a way that the low-energy spectrum consists of two
 singlets and one triplet. The symmetry of the
pertinent quantum states is described by a non-compact
 group, that is,  $SO(5)$. Decoherence and dephasing will be
 studied through the application of a time-dependent gate voltage
 to the side dot in a TDQD, and its influence on the resonance Kondo
tunneling. Usually, decoherence in Kondo tunneling arises due to
non-equilibrium spin flip processes induced by an external
time-dependent potential. \cite{KNG} Here we propose an alternative
mechanism, which involves processes with charge transfer between
the two wells of the TDQD. As a result, the singlet excitons $(E)$ are
involved in dephasing and decoherence processes. It will be shown
that these processes arise due to dynamical gauge
fluctuations, which accompany singlet-triplet transitions within
various multiplets of the $SO(5)$ group.

The structure of the paper is as follows: The underlying
time-dependent Hamiltonian is introduced in section II, first as a
generalized Anderson impurity model, and then in terms of Hubbard
operators. Section III is devoted to the derivation of an
effective spin Hamiltonian (following an appropriate
Schrieffer-Wolff transformation) and presentation of {\it time
dependent} scaling equations for the relevant exchange constants.
Fluctuation corrections to the scaling equations are thoroughly
discussed in section IV. This section
 is the central one, as it exposes the physical origin of dephasing, decoherence
and relaxation in the present model system. Since it might appear
too technical, it is following by
concluding remarks in section V, where the main results are
explained verbally. Related mathematical topics such as
necessary ingredients of Group
Theory,  manipulations of Vertex Corrections and
some aspects of analytical continuation as used  in the calculations are
explained in the following three Appendices.

\section{Model and time-dependent Hamiltonian}

\begin{figure}[h]
\includegraphics[width=6cm,angle=0]{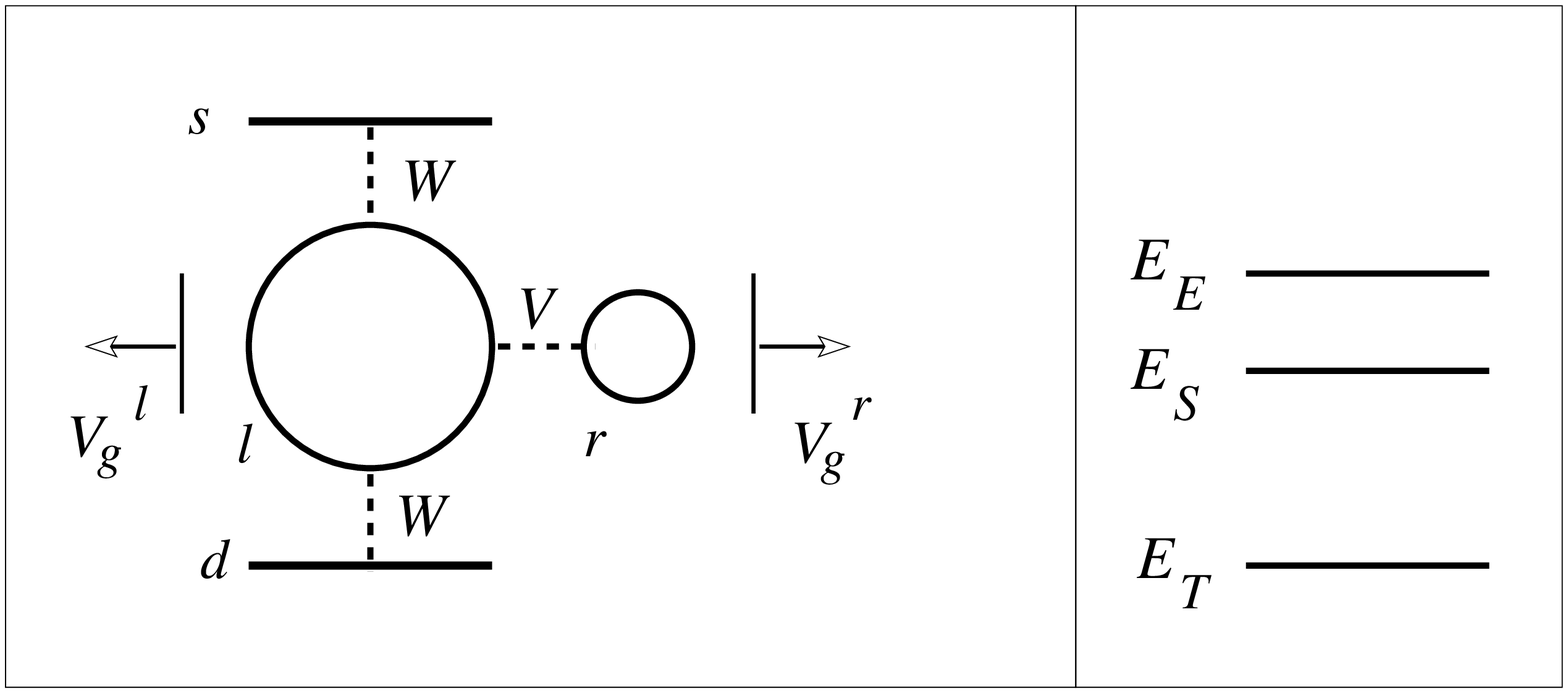}
\caption{Left panel: T-shaped double quantum dot (TDQD); right
panel: Ground state and lowest excitations of doubly occupied TDQD
renormalized by interaction with leads.}\label{f.00}
\end{figure}

We consider an asymmetric DQD studied in Ref. \onlinecite{KA01},
where only the left dot is coupled to the leads (Fig. \ref{f.00},
left panel).
 The capacitive
energies in the left and right dots are different, $Q_r\gg Q_l$ so
that the Coulomb blockade completely suppresses doubly occupied
states in the right dot. The gate voltages $V_g^{l,r}$ are applied
separately to the left and right dots.
The new ingredient here is the introduction of a
 small "trembling" (time dependent) component  in the
gate voltage of the right dot, so that
$$V_g^r= V_g^r(0) + v_g(t).$$

Let us first recall the spectrum of the TDQD in the absence
of the time dependent component.
The constant parts of the gate voltages are included in the energy
levels of the double dot,
$\epsilon_{l,r}=\varepsilon_{l,r}+V_g^{l,r}(0)$. These voltages
are tuned in such a way that
$$
\epsilon_l+\epsilon_r < 2\epsilon_l+ Q_l\ll 2\epsilon_r+Q_r.
$$
Hereafter,  the Fermi energy in the leads is chosen as the zero energy
level.
As excited states of the TDQD with two electrons in the right
dot are excluded from the low-energy part of the spectrum,
 the low energy levels of the TDQD are \cite{KA01}
\begin{eqnarray}
&& E_T = \epsilon_l +\epsilon_r - M_T, \label{dotnot} \\
&& E_S =  \epsilon_l+\epsilon_r  -2\beta V - M_S, \nonumber \\
&& E_E = 2\epsilon_l+ 2\beta V +M_E, \nonumber
\end{eqnarray}
where $V$ is the potential which acts between the dots, see Fig.1.
The notations $T, S, E$ respectively refer to triplet, singlet and exciton states of the
TDQD. In the exciton state, the two electrons reside on the left dot.
The above results are obtained in first order
of the small parameter  $\beta=V/(\varepsilon_l-\varepsilon_r+ Q_l)\ll 1$.
Anticipating further analysis, we include in the low energy spectrum
 the level
shifts $M_{\Lambda=T,S,E}$ which result from renormalization of
the dot levels due to the tunnel contact with the leads (the so called
"Haldane renormalization" \cite{Hald}, see below). It was shown in
Ref. \onlinecite{KA01} that $M_T > M_S$, so that the singlet and
triplet levels may cross due to this renormalization. Such level
crossing occurs provided the charge transfer exciton energy $E_E$ is
not very high. Here we study just this regime, so that
taking into account the above level renormalization, the ground state of
the TDQD in a contact with the leads is a triplet state $E_T$ ( Fig.
\ref{f.00}, right panel).

 Having in mind the above
energy level scheme (\ref{dotnot}) for the TDQD with {\it constant} gate
voltages $V_g^{l,r}(0)$, we now consider the influence of the
trembling potential $v_g(t)$ on the dot spectrum. As usual, one
should discriminate between slow and fast components of a temporal
perturbation and treat the former processes in an adiabatic
approximation. In order to separate adiabatic and non-adiabatic
parts of the trembling potential it is useful to introduce the spectral
density $A(\Omega)$ of the trembling potential $v_g$:
\begin{equation}\label{tremb}
v_g (t) =v_g\int d\Omega A(\Omega) e^{i\Omega t}.
\end{equation}
In the present study we are mainly interested in the nearly adiabatic
regime, where the spectral density is concentrated in the
frequency interval $0<\Omega<\Omega_a$ with $\Omega_a\ll
\Delta_{ST} = E_S - E_T$. In this case the contribution of the
trembling potential may be treated by means of time-dependent
scaling approach, while non-adiabatic corrections should be
considered perturbatively.

The Hamiltonian describing the system schematically shown in Fig.
\ref{f.00} has the following form:
\begin{equation}\label{1.0}
H= H_{band} + H_{dot} + H_{tun}.
\end{equation}
Here the first term is the band Hamiltonian, which describes
electrons in the leads [source ($s$) and drain ($d$),
respectively],
\begin{equation}
H_{band}=\sum_{b=s,d}\varepsilon_{kb}
c^\dag_{kb\sigma}c_{kb\sigma}~,
\end{equation}
$k,\sigma$ are the wave vector and spin projection, respectively.

The tunneling Hamiltonian involves only electrons in the left dot.
\begin{equation}\label{htun}
H_{tun}=W\sum_{k\sigma}( c^\dagger_{k\sigma}d_{l\sigma}+ H.c.),
\end{equation}
where the operators $d_{l\sigma}$ correspond to electrons in the
TDQD. In the tunneling part \ref{htun}
we have already excluded one of the two channels from the tunneling
Hamiltonian by introducing even and odd combinations of electron
wave function in two leads. Then, only the even standing wave
$c_{k\sigma}=\frac{1}{\sqrt{2}}(c_{ks\sigma }+c_{kd\sigma})$
enters $H_{tun}(0)$ (we consider a completely symmetric device
with lead-dot tunneling constants $W_s=W_d=W$).

It is useful at this point to carry out some manipulations
which will turn the discussion more transparent.
In the first step, a canonical
transformation is performed, \cite{BRUS,GA,KNG} which
eliminates the trembling potential from the diagonal part of the
dot Hamiltonian
\begin{equation}
H_{dot}=H_{l0}+ H_{lr0} + H_r ~, \label{1.1}
\end{equation}
with
$$
H_{l0}=\varepsilon_l n_l +Q_ln_l^2,$$
$$
H_{lr0}=V\sum_{\sigma}(d_{l\sigma}^\dagger d_{r\sigma} + H.c.),$$
$$H_r=\varepsilon_r n_r +Q_rn^2_r +V_g(0)n_r +v_g(t)n_r =
H_{r0}+H_r(t).
$$
Here $n_{j}=d^\dag_{j\sigma}d_{j\sigma}$ is the density operator
for the electrons in the left $(j=l))$ and right ($j=r$) dot.
The required transformation is,
\begin{equation}\label{1.4}
\widetilde{H}=U_1HU_1^{-1}-i\hbar \frac{\partial U_1}{\partial
t}U_1^\dag,
\end{equation}
with $U_1 \equiv \exp[-i\Phi_1(t)n_r] $ where the phase
$\Phi_1(t)$ is given by
$$
\Phi_1(t)=\frac{1}{\hbar} \int^tdt^\prime v_g(t^\prime).
$$
This phase may be rewritten in terms of the spectral density
introduced in Eq. \ref{tremb}:
\begin{equation}
\Phi_1(t)=\int d\Omega \Phi_1(\Omega)e^{i\Omega t} \equiv  v_g\int
d\Omega\frac{A(\Omega)}{(i\hbar\Omega)}e^{i\Omega t}. \label{1.9}
\end{equation}
Then, expanding the exponent in terms of the weak trembling
potential, we come to the expression,
\begin{eqnarray}
\tilde{H}_{dot} &=& H_{dot}^{(0)}+iV\Phi_1(t)
\sum_{\sigma}(d^\dagger_{r \sigma} d_{l \sigma}-H.c.)\nonumber
\\
& \equiv & H_{dot}^{(0)}+H_{dot}(t)~. \label{1.2}
\end{eqnarray}

In the next step, it proves to be more convenient to work in a representation
$|\Lambda\rangle$, which diagonalizes the dot Hamiltonian, i.e. to
rewrite it in terms of Hubbard operators
$X^{\Lambda\Lambda'}=|\Lambda\rangle\langle\Lambda'|$. After the
standard diagonalization procedure for the {\it time independent}
part of the Hamiltonian we have \cite{KA01}
\begin{equation}
H_{dot}^{(0)}=\sum_\Lambda E_\Lambda
X^{\Lambda\Lambda},~~~~~~~~\Lambda=T\mu,S,E. \label{1.7}
\end{equation}
Recall again that $T,S,E$ stand for triplet, singlet and charge transfer
exciton states of TDQD, respectively while $\mu=\pm1,0$ are the spin
projections of the triplet state. The diagonal Hubbard operators are constrained by the
condition
\begin{equation}\label{constr}
\sum_\Lambda X^{\Lambda\Lambda}=1.
\end{equation}

In terms of Hubbard
operators the tunneling Hamiltonian (\ref{htun}) acquires the form
\begin{equation}\label{htun1}
H_{tun}^{(0)}=W\sum_{\Lambda,
\sigma\sigma'}\left(c^\dag_{k\sigma}X^{r_{\sigma'}\Lambda} + H.c.
\right).
\end{equation}
Here the index $r_{\sigma'}$ stands for the electron in the right
well with spin projection $\sigma'$, which remains in the TDQD,
when the electron $l_\sigma$ escapes  from the left well to the
lead. As was mentioned above, the lead state $k\sigma$ means the
even combination of lead electrons with the wave vector $k$.  All
tunneling transitions are described by the same constant $W$.

Writing $H_{dot}(t)$ from Eq. (\ref{1.2}) in the form $
H_{dot}(t)=iV\Phi_1(t){\cal S}_{lr} $ where
$$
{\cal S}_{lr}=\sum_{\sigma}(d^\dagger_{r \sigma} d_{l
\sigma}-H.c.),
$$
one finds
\begin{equation}
{\cal S}_{lr}|S\rangle = \sqrt{2}|E\rangle, ~~~ {\cal
S}_{lr}|E\rangle = -\sqrt{2}|S\rangle, ~~~ {\cal
S}_{lr}|T\rangle=0. \label{1.3}
\end{equation}
It follows from these equations that
\begin{equation}\label{1.33}
{\cal S}_{lr}=\sqrt{2}(X^{ES}-X^{SE})\equiv iA\sqrt{2}.
\end{equation}
The new scalar operator $A$ is one of the generators of the
$SO(5)$ group, which characterizes the dynamical symmetry of a
biased DQD (see Appendix A).
 Then, the commutation with the
diagonal Hubbard operators yields
\begin{equation}
[{\cal S}_{lr},X^{SS}]=iA\sqrt{2},~~~ [{\cal
S}_{lr},X^{EE}]=-iA\sqrt{2} \label{anti}
\end{equation}
(${\cal S}_{lr}$ is an anti-Hermitian operator).
As a result, the  GA transformation \cite{GA} leads to
\begin{equation}
H_{dot}(t) = -\sqrt{2}V\Phi_1(t)A. \label{1.5}
\end{equation}

\section{Time-dependent poor man's scaling}

Next, the Haldane-Anderson scaling approach \cite{Hald} should be
applied to $H_{dot}+H_{band}+H_{tun}$ in order to rescale the
energies in the TDQD. Here $H_{dot}$ (\ref{1.2}) contains the
time-dependent component. Unlike the static case considered in
Ref. \onlinecite{KA01}, we deal here with a TDQD whose
Hamiltonian is non-diagonal in
$\Lambda$, such that the off-diagonal elements
are time-dependent. However,
for the slow trembling potential with characteristic frequencies
$\Omega_a \ll \Delta_{ES} = E_{E}-E_{S}$, one may treat this
time-dependent term {\it adiabatically} at least in zero-order
approximation.
This means that before turning to the Haldane RG procedure, one
has to get rid of the non-diagonal $SE$ mixing terms with the help
of a second time-dependent canonical transformation. This
transformation is given by the matrix $U_2=\exp \Theta$, which is
found from the condition  \cite{KNG}
\begin{equation}\label{1.11}
H_{dot}(t)+[\Theta,H_{dot}^{(0)}]=i\hbar\frac{\partial
\Theta}{\partial t}.
\end{equation}
Within the adiabatic approximation, one may carry out this
diagonalization procedure at
each moment $t$ neglecting any retardation and omitting the term
in the r.h.s. of this equation. Then the matrix $\Theta$ is given by
\begin{equation}\label{u2}
\Theta= \frac{i\sqrt{2}V\Phi_1(t)}{\Delta_{ES}}(X^{ES}+X^{SE}).
\end{equation}

At this stage, we apply the "adiabatic" Haldane RG procedure, i.e.
calculate the scaling trajectories of the energy states, which
evolve with the reduction of the energy scale in the metallic
reservoir from the initial value $D_0$ to the actual value $D$.
Since the transformations (\ref{1.4}) and (\ref{1.11}) do not
involve the triplet state, the scaling trajectory $E_{T}$ as a
function of scaling variable $\xi=\ln (D_0/D)$ is the same as in
our previous calculations. \cite{KA01} As a result, the triplet
state includes only the time-independent Haldane shift,
 whereas the two singlet states acquire time dependence,
\begin{eqnarray}
E_T &\to & E_T -M_T(\xi) , \label{adia}\\
E_S(t)&\to& E_S- M_S(\xi) - C_S\Phi_1^2(t), \nonumber \\
E_E(t)&\to& E_E+ M_E(\xi) + C_S\Phi_1^2(t). \nonumber
\end{eqnarray}
Here
 $C_S={2V^2}/\Delta_{ES}$ is an additional time-dependent adiabatic
 part within the Haldane renormalization scheme.
Thus, the term $H_{dot}^{(0)}$ in the dot Hamiltonian  (\ref{1.2}) is
given by Eq. (\ref{1.7}) and
\begin{equation}\label{hdt}
H_{dot}(t)=C_S\Phi_1^2(t)(X^{EE}-X^{SS}).
\end{equation}
There is no correction  to the matrix $\Theta$ from the time
derivative in the
 r.h.s. of Eq. (\ref{1.11}) at least up to first order in
the adiabatic parameter
 $\kappa_1=v_g(t)/\Delta_{ES}$.
Besides, due to the same transformation $U_2$,  the components of
the tunnel Hamiltonian containing operators $X^{E\lambda}$ and
$X^{S\lambda}$ become time dependent as well,
\begin{equation}\label{ht}
H_{tun}=H_{tun}^{(0)}+H_{tun}(t).
\end{equation}
The "trembling tunneling" contribution in these terms has the form
\begin{equation}
H_{tun}(t)=-i\sqrt{2}\frac{VW\Phi_1(t)}{\Delta_{ES}}
\sum_{k\sigma}(X^{Sr_{\bar{\sigma}}} + X^{Er_{\bar{\sigma}}})
c_{k\sigma}-H.c. \label{1.6}
\end{equation}
Thus, the trembling gate voltage generates time-dependent
tunneling through the singlet states, but the operator form of
$H_{tun}$ is conserved.

We see that the trembling potential involves a contribution of the
excited singlet in the tunneling Hamiltonian, and the latter
introduces fluctuations in tunneling through the ground state
singlet.  The spin $S=1$ states are not involved in these
trembling processes. However, eventually we expect that the
triplet is subject to dephasing  via the operators ${\bf P}$ and
${\bf M}$ (see equation \ref{SW2} below)
due to dynamical $SO(5)$ symmetry of the TDQD (see
Appendix A). One should remember that the state $|E\rangle$ enters
the manifold of eigenstates of the isolated TDQD (\ref{1.7}). In
the static case this high energy {\it charge} excitation is
admixed with the low-lying singlet {\it spin} state $|S\rangle$.
This admixture changes the lead-dot tunneling rate  and can lead
to the inequality $M_T>M_S$ mentioned earlier. We consider the
situation well beyond the S/T crossover so that the positive
singlet-triplet energy gap satisfies the inequality
$\Delta_{ST}=E_S-E_T \gg \Omega_a$. The adiabatic Haldane
procedure (\ref{adia}) results in a time-dependent S/T energy gap
\begin{eqnarray}
&&\Delta_{ST}(t) = E_S-E_T - \delta (t), \nonumber\\
&&\delta (t)= C_S\Phi_1^2(t). \label{1.8}
\end{eqnarray}

The adiabatic contribution of a charge transfer exciton $(E)$ to
the effective indirect exchange (cotunneling) Hamiltonian may be
taken into account by means of a time dependent Schrieffer-Wolff
transformation, which includes $H_{tun}(t)$ (\ref{1.6}). This
transformation is described in Ref. \onlinecite{KNG} for the dot
occupied by a single electron with $S=1/2$. In our case the
exchange Hamiltonian is affected by the trembling perturbation in
spite of the fact that the localized spin ${\bf S}=1$ is not
directly affected by the time dependent potential. However, charge
fluctuations induced by time dependent tunneling (\ref{1.6})
perturb the $S$- and $E$-states, which are connected with the spin
by the kinematics of vector and scalar operators of the dynamical
symmetry group $SO(5)$.  All these corrections can be incorporated
in the Schrieffer-Wolff (SW) Hamiltonian $H_{SW}$ for the dot
obeying the SO(5) symmetry. The time-independent part of the SW
Hamiltonian takes the form \cite{KKA}
\begin{eqnarray}
H_{SW}^{(0)}& = & H_{dot}^{(0)} + H_{cotun}^{(0)},\label{SW1} \\
H_{dot}^{(0)}& = & \frac{1}{2}\left(E_T {\bf S}^2 + E_S {\bf P}^2+
E_E {\bf M}^2 \right) +  Q({\hat N}-2)^2, \nonumber \\
H_{cotun}^{(0)} &=& J^T_0 {\bf S}\cdot {\bf s} + J^{ST}_0 {\bf
P}\cdot {\bf s}+J^{ET}_0 {\bf M}\cdot {\bf s}.\nonumber
\end{eqnarray}
Here the terms describing the irrelevant potential scattering are
omitted (see, however, the discussion of fluctuation corrections
in Section IV).

 The Hamiltonian  (\ref{SW1}) is expressed in terms of SO(5) group
generators presented in Appendix A.
 Here $\hat N$ is the operator of particle
occupation number, and the last term in (\ref{SW1}) describes the
constraint $N=2$ imposed by the Coulomb blockade $Q$ on the charge
sector of the Hilbert space. As we mentioned already, the operator
${\bf S}$ is not affected by the time-dependent canonical
transformations. However, the operators ${\bf P}$ and ${\bf M}$
 will be involved in the
time-dependent processes.

We generalize the adiabatic SW transformation derived in Ref.
\onlinecite{KNG} by introducing the matrix $U_3=\exp i\Upsilon$
with $\Upsilon$ defined as
\begin{equation}\label{2.1}
\Upsilon(t)=\sum_{k\sigma}\left[\upsilon_k^S(t)X^{Sr_{\bar{\sigma}}}
c_{k\sigma}+ \upsilon_k^E(t)X^{Er_{\bar{\sigma}}} c_{k\sigma} -
H.c.\right].
\end{equation}
The coefficients $\upsilon_k^\Lambda$ are chosen to eliminate the
time dependent terms (\ref{1.6}) in the effective Hamiltonian
\begin{equation}
H_{SW} = \exp(-i\Upsilon)[H_{dot}+ H_{band} +H_{tun}]\exp
i\Upsilon,
\end{equation}
where the first and third terms in the Hamiltonian depend on time.
To fulfil this request, the transformation matrix should satisfy
the equation
\begin{equation}\label{1.12}
H_{tun}+ [\Upsilon, (H_{dot}+ H_{band})] = i\hbar\frac{\partial
\Upsilon}{\partial t}
\end{equation}
(cf. Eq. \ref{1.11}). Repeating the procedure used in the
calculation of the matrix $U_2$ and including the first order
corrections in the adiabaticity parameters $\kappa_2^{S}(t)=
v_g(t)/\epsilon_l$ and $\kappa_2^{E}(t)= v_g(t)/(2\epsilon_l  -
\epsilon_r)$
 from the time derivative on the
r.h.s. of Eq. (\ref{1.12}), one finds the following expression for
$\Upsilon(t)$
\begin{eqnarray}
&&\Upsilon(t)=\Upsilon_T+\Upsilon_S(t)+\Upsilon_E(t),
\label{1.13}\\
&&\Upsilon_T=\frac{W}{\epsilon_l-M_T}\sum_{k\sigma\sigma'}
\sum_{\nu=0,1,\bar1}\left(
X^{\nu,r\sigma'}c_{k\sigma}- H.c. \right),\nonumber\\
&&\Upsilon_S(t)=\frac{W-w_2^{S}(t)}
{\epsilon_l-M_S(t)}\sum_{k\sigma}(X^{S,r_{\bar\sigma}}
c_{k\sigma}- H.c.),\nonumber\\
&&\Upsilon_E(t)=\frac{W-w_2^{E}(t)}
{2\epsilon_l-\epsilon_r+M_E(t)}\sum_{k\sigma}(X^{E,r_{\bar\sigma}}
c_{k\sigma}- H.c.).\nonumber
\end{eqnarray}
 The time-dependent quantities in $\Upsilon_S$ and
$\Upsilon_E$ are
$$
M_S(t)= M_S - C_S\Phi_1^2(t),\;\;\; M_E(t)= M_E + C_S\Phi_1^2(t)
$$
$$
w_2^{S}(t)=\frac{VW\sqrt{2}\Phi_2^{S}(t)}{\Delta_{ES}},\;\;\;
w_2^{E}(t)=\frac{VW\sqrt{2}\Phi_2^{E}(t)}{\Delta_{ES}},
$$
with $\Phi_2^{S}(t)=\Phi_1(t)-\alpha_2^{S}(t)$ and
$\Phi_2^{E}(t)=\Phi_1(t)-\alpha_2^{E}(t)$.

This procedure results in the appearance of a time-dependent
component in $H_{cotun}$, which modifies the coupling constants
$J^{ST}$ and$J^{ET}$ in the Hamiltonian (\ref{SW1}). As a result,
we get the following effective exchange Hamiltonian which induces
the decoherence and dephasing effects in the Kondo tunneling
through TDQD:
\begin{eqnarray}
H_{cotun} &=& J^{TT}_0 {\bf S}\cdot {\bf s} + J^{ST}(t) {\bf
P}\cdot {\bf s}+J^{ET}(t) {\bf M}\cdot
{\bf s}.\nonumber \\
\label{SW2}
\end{eqnarray}
Here
\begin{equation}\label{2.3}
J^{ST}(t)=\frac{W-w_2^{S}(t)}{\epsilon_l-M_S(t)}
\cdot\frac{W(1-V/\Delta_{ES})\sqrt{2}\Phi_1(t)}{\Delta_{ES}},
\end{equation}
\begin{equation}\label{2.3}
J^{ET}(t)=\frac{W-w_2^{E}(t)}{2\epsilon_r-\epsilon_l+M_E(t)}
\cdot\frac{W(1-V/\Delta_{ES})\sqrt{2}\Phi_1(t)}{\Delta_{ES}},
\end{equation}
and the time-dependent parts of the coupling constant $J^{ST}$ and
$J^{ET}$ may be obtained by expansion of its general form in the
adiabatic parameters $\kappa_{1},\kappa_{2}^{S},\kappa_{2}^{E}$.

The first manifestation of this time dependence, which can be seen
already within an adiabatic approximation is the uncertainty in
the definition of the Kondo temperature. To describe this effect,
we refer to the poor man's scaling equations for the Kondo effect
in DQD for time-independent Hamiltonian (\ref{SW1}) derived
previously. \cite{KA01} In the adiabatic regime, these equations
retain their form, but the coupling constants depend on time $t$
as a parameter. In particular in the limit when the exciton state
$|E\rangle$ is quenched, the system of scaling equations has the
form
\begin{eqnarray}\label{scale}
dj_1/d \xi  = - j_1^2-[j_2(t)]^2;~~~ dj_2 /d \xi = - 2j_1j_2(t).
\end{eqnarray}
Here $j_1=\rho_0J^{TT}, ~j_2=\rho_0J^{ST}, ~j_3=\rho_0J^{ET}$,
$\rho_0$ is the density of states at the Fermi level.

As is already established in the theory of Kondo-effect at a
singlet-triplet (S/T) crossing,\cite{KA01,GT01,Pust1,Eto} in the
limit $\Delta_{ST}\gg T_{K}$ the solution of these equations may
be expressed in terms of $T_{K0}=D_0 \exp[-1/(j_1+j_2)]$:
\begin{equation}\label{asymp}
\frac{T_K}{T_{K0}}=\left[\frac{T_{K0}}{\Delta_{ST}(t)}
\right]^\zeta,
\end{equation}
where $\zeta \lesssim 1$ is a constant. The relative amplitude
$\delta T_K$ of adiabatic variations of the "time-dependent Kondo
temperature" \cite{KNG} may be estimated from this equation:
\begin{equation}
\delta T_K \approx \left[\frac{T_{K0}}{\Delta_{ST}(0)}
\right]^{1+\zeta}  2C_S \Phi_1^2
\end{equation}
($\Phi_1^2$ is the mean square value of the trembling parameter).
In this asymptotic region, the time-dependent corrections are
 insignificant. In accordance with the above mentioned theory of an S/T
 transition for a time-independent
 gate voltage,\cite{KA01,GT01,Pust1,Eto} the
 Kondo temperature increases with decreasing $\Delta_{ST}$ and
 reaches its maximum $T_K = T_{K0}$ at a crossing point
 $\Delta_{ST}=0$. One may expect that the role of non-adiabatic
 corrections increases with decreasing $\Delta_{ST}$. As a result,
 the scaling behavior should be seriously violated when approaching the
 crossing point, and the deviations
 from the prescriptions of an adiabatic theory in the dependence
 $T_K(\Delta_{ST})$ should grow accordingly.

\section{Fluctuation  corrections to RG equations}

In this section we go {\it beyond} the adiabatic approximation and
take into account decoherence and dephasing corrections to the
Kondo tunneling. Unlike the mechanisms described in Ref.
\onlinecite{KNG}, where the time-dependent spin-flip processes
were the source of dephasing, we appeal here to gauge
fluctuations, which arise because {\it two} singlet states
$|S\rangle$  and $|E\rangle$ are involved in the formation of a
Kondo resonance in a triplet ground state $|T\rangle$ in the
process of time-dependent cotunneling.

The first source of non-adiabatic corrections is associated with
the fluctuations of the energy gap $\Delta_{ST}$ (\ref{1.8}),
which may be converted into  gauge fluctuations of a Casimir
operator (\ref{A.4}). Using an obvious chain of subsets,
$$U(1)\subset SU(2) \subset SO(4) \subset SO(5),$$
we conclude that the trembling potential affects the least
continuous subgroup of $SO(5)$.

To perform this conversion we turn to the fermionic representation
of the generators of the $SO(5)$ group (see Appendix A). We now
work in the subspace of low-energy states $|T\rangle, |S\rangle$,
which are described by the dynamical symmetry group $SO(4)$. Then
the non-adiabatic fluctuations of the energy gap $\Delta_{ES}$ may
be described as fluctuations of the "reduced" Casimir operator
${\bf S}^2+ {\bf P}^2$, which, in turn is rewritten as a local
constraint (\ref{const2}) where the last term is eliminated. We
introduce the "fluctuating constraint" in $H_{dot}$ by means of a
{\it time-dependent Lagrange factor $\mu$:}
\begin{equation}\label{1.10}
H_{dot}^\prime=H_{dot}(0) - \mu(t)(\hat{R} -1)
\end{equation}
with
$$
\hat{R}= f^\dagger_s f_s +\sum_{\nu=1,0, \bar 1}f^\dagger_\nu
f_\nu~.
$$
Thus, the fluctuating constraint takes into account time-dependent
admixture of the $E$-state to the $S$-state. Although the exciton
$E$ is excluded from the effective spin space, its effects are
present through a "randomization" of the constraint. The
fluctuating time-dependent Lagrange factor can be eliminated from
the $SO(4)$ fermionic Green's functions by means of the {\it
global} $U(1)$ gauge transformation performed in all sectors of
the $f$-representation
\begin{equation}\label{4.1}
f_\lambda \to f_\lambda e^{i\phi(t)}.
\end{equation}
Minimizing the free fermion part of the Lagrangian, one has $
\hbar \dot \phi(t)=\mu(t)$ (here $\lambda=\bar1,0,1,s$).

On the other hand, the constraint fluctuations in terms of
time-dependent Lagrange factors can be reformulated as a
propagation of fermions along the time axis in the presence of a
time-dependent external scalar potential. The solution of this
problem results in the appearance of an imaginary part in the zero
frequency spin fermion propagators, ${\rm Im}\Sigma_f(0)=\gamma$
(see below). Eventually, the existence of this damping prevents
the formation of a coherent Kondo tunneling regime through DQD at
zero temperature, so the effects connected with this type of gauge
fluctuations can be qualified as {\it decoherence}.

Another source of non-adiabatic gauge fluctuations comes from  the
time-dependent poor man's scaling solution. The co-tunneling
Hamiltonian (\ref{2.3}) contains the time-dependent corrections to
${\bf P \cdot s}$ and ${\bf M\cdot s}$. In this case (unlike the
first mechanism), the source of gauge fluctuations is due to
non-diagonal operators {\bf P}, {\bf M}, so that the
time-dependent coupling constants are parametrized as
$j_2(t)=j_2e^{i\theta_{ST}(t)}$ and
$j_3(t)=j_3e^{i\theta_{ET}(t)}$.

The phases in $j^{\Lambda\Lambda'}$ can be eliminated by the {\it
local} $U(1)$ gauge transformation
\begin{equation}\label{gau}
f_\lambda \to f_\lambda e^{i\vartheta_{\lambda}(t)}
\end{equation}
performed in the $S$-, $T$- and $E$-sectors of the fermionic
representation of the
 $SO(5)$ group. The local gauge phases may be represented as
\begin{equation}\label{phasis}
\theta_{ST}(t)=\vartheta_T(t)-\vartheta_S(t),\;\;
\theta_{ET}(t)=\vartheta_T(t)-\vartheta_E(t)
\end{equation}
so that the phases $\vartheta_\lambda(t)$ have different time
derivatives. Therefore, the effects related to this type of gauge
fluctuations are, in fact, {\it dephasing} effects stemming out of
weakly non-adiabatic $TS$ and $TE$ transitions.

\begin{figure}[h]
\includegraphics[width=4cm,angle=0]{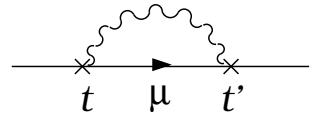}
\caption{Self energy correction due to gauge fluctuations in
spin-fermion propagator.}\label{f.0}
\end{figure}

Now we turn to the calculation of gauge fluctuation corrections in
the self energies and vertices entering the Kondo tunneling
diagrams. To calculate the damping of retarded spin-fermion
propagators in the triplet sector of phase space
\begin{equation}\label{get}
G_{T\mu}(t-t')=\langle
f_\mu(t)f^\dag_\mu(t')\rangle_R=-i\langle[f_\mu(t)f^\dag_\mu(t')]_+\rangle
\end{equation}
we apply perturbation theory for time-dependent external
potentials to the last term in the Hamiltonian (\ref{1.10}). In
the corresponding diagrammatic technique the times $t$ are marked
by crosses and the correlation functions connecting the
fluctuations related to the instants $t,t'$ are denoted by wavy
lines. Then the first order diagram for the self energy of
$G_{T\mu}(t-t')$ is given by the diagram Fig. \ref{f.0} (the
spin-fermion propagator is represented by the full line). We
consider the simplest case of white-noise correlations
$\hbar^2\langle\dot\phi_1(t)\dot\phi_1(t')\rangle=r_0\delta(t-t').$
The corresponding contribution to the self energy is a purely
imaginary damping
\begin{equation}\label{tau}
\gamma_1 =\hbar/2\tau_d\sim C_S^2 r_0.
\end{equation}
This first order approximation is valid as long as the damping is
weak in comparison with the Kondo temperature, $\hbar /\tau_d\ll
T_K$. If these two quantities are comparable then the whole
sequence of "rainbow" diagrams should be inserted in the
spin-fermion self energy. This means that instead of the white
noise we get a more realistic description of trembling, which
takes  the retardation effects (non-gaussian corrections) into
account. This specification is not too essential for our purposes.
More important is the low-energy cutoff  $\sim \hbar /\tau_d$ in
the RG procedure, which prevents crossover to the strong-coupling
limit of the Kondo tunneling due to incoherent phase fluctuations.

Our next goal is to find non-adiabatic corrections for exchange
vertices, which contribute to the dephasing mechanism mentioned
above (see Eq. \ref{phasis}). Dephasing means that the gauge
transformation (\ref{gau}) eliminates the phase $\theta_{ST}$
within the accuracy of phase fluctuations, i.e. the local gauge
phases have the form
\begin{eqnarray}
\theta_{ST}(t)=\vartheta_T(t)-\vartheta_S(t)+\varphi_s(t),\nonumber \\
\theta_{ET}(t)=\vartheta_T(t)-\vartheta_E(t)+\varphi_e(t).\label{phasis1}
\end{eqnarray}
Then, expanding the exponents in the non-adiabatic vertex
corrections $J^{ST}\left[e^{i\varphi_s(t)}-1\right]$ and
$J^{ET}\left[e^{i\varphi_e(t)} -1\right]$, we obtain the
fluctuating part of the Hamiltonian (\ref{SW2}) in a form
\begin{eqnarray}\label{fluflu}
\delta H_{cotun}= i \left[ J^{ST}\varphi_s(t){\bf P}\cdot {\bf s}
 + J^{ET}\varphi_e(t) {\bf M}\cdot {\bf s}\right].
\end{eqnarray}

\begin{figure}[h]
\includegraphics[width=7cm,angle=0]{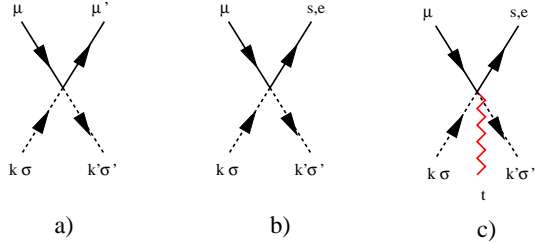}
\caption{(a) Adiabatic exchange vertices for triplet channel $T$;
    ( b) Similar diagrams for $ST$ and $ET$ channels;
    (c) non-adiabatic corrections to two latter vertices.}\label{f.1}
\end{figure}

Adiabatic vertices (\ref{SW1}) and non-adiabatic corrections
(\ref{fluflu})  may be presented in a graphical form (Fig.
\ref{f.1}). The bare interactions are shown in Figs.
\ref{f.1}(a,b) by means of four-tail vertices where the operators
${\bf S}$, ${\bf P}$ and ${\bf M}$ are expressed via fermion
operators in accordance with Eqs. (\ref{fff}). Here solid lines
stand for spin fermions and dashed lines represent conduction
electrons.

The time-dependent vertices  (\ref{fluflu}) are shown in Fig.
\ref{f.1}c, where the broken line stands  for the fluctuation
field $D(t)= \langle \varphi(t)\varphi(0)\rangle_R$.

If the fluctuating trembling signal is characterized by
retardation,
\begin{equation}\label{demp}
D(t)= -i\alpha^2 e^{-\zeta|t|}
\end{equation}
then the Fourier transform of this correlation function is
\begin{equation}\label{cor}
D(\omega)=\frac{2i\alpha^2\zeta}{\omega^2+\zeta^2 }.
\end{equation}
Other elements entering the vertices are $G_{T\nu}(t-t')$ defined
in Eq. (\ref{get}), a similar propagator for spinless fermions
representing the singlet state
\begin{equation}\label{ges}
G_S(t-t')=\langle f_s(t)f^\dag_s(t')\rangle_R .
\end{equation}
and conduction electron propagators
\begin{equation}\label{gek}
g_{k\sigma}(t-t')=\langle
c_{k\sigma}(t)c^\dag_{k\sigma}(t')\rangle_R .
\end{equation}
The Fourier transforms of the Green functions (\ref{get}),
(\ref{ges}) and (\ref{gek}) are
\begin{eqnarray}
G_{T\mu}(\omega)&=&(\omega+i\eta)^{-1},~~G_{S}(\omega)=(\omega-\Delta_{TS}+i\eta)^{-1},\nonumber
\\
&&g_{k\sigma}(\omega)=(\omega-\varepsilon_{k\sigma}+i\eta)^{-1}.
\label{gef}
\end{eqnarray}
(here $\eta \to +0$).

\begin{figure}[h]
\includegraphics[width=8cm,angle=0]{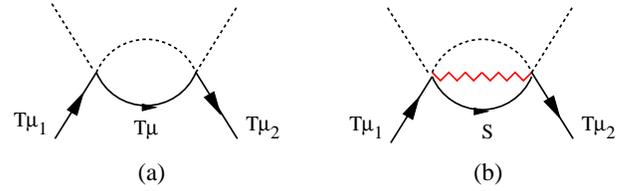}
\caption{First leading parquet diagram (a) and first non-adiabatic
correction to it (b). The pseudofermion, electron and fluctuation
propagators are represented by solid, dotted and broken lines,
respectively. The absence of direction in electron propagator
lines assumes that both directions (clockwise and anti-clockwise)
are  possible.}\label{f.2a}
\end{figure}

\begin{figure}[h]
\includegraphics[width=4cm,angle=0]{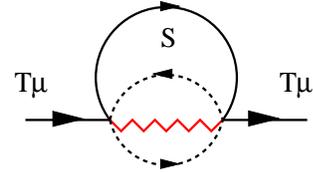}
\caption{First non-adiabatic correction to propagator
$G_{T\mu}$.}\label{f.2b}
\end{figure}

 The first fluctuation correction to adiabatic vertices Fig.
\ref{f.2a}a is displayed in Fig. \ref{f.2a}b.
Although the contribution of the diagram in Fig. \ref{f.2a}b is
parametrically small compared with that of Fig. \ref{f.2a}a due to
a small factor $(\alpha j_0^{ST})^2 \ll 1$, this diagram
represents a building block for construction of non-adiabatic
corrections for vertices and self-energy parts. These corrections
affect both adiabatic vertices (see Appendix B) and the self
energy $\Sigma_{T}(\omega)$ of the spin-fermion propagator
$G_{T\mu}$ (Fig. \ref{f.2b}). The latter diagram may be obtained
from the vertex (Fig.\ref{f.2a}b) by gluing two free electron
lines in the electron propagator. Besides, the diagram (Fig.
\ref{f.2b}) is connected with first non-parquet TT vertex by a
Ward identity (see Appendix B). We will use this identity below,
when calculating the spin relaxation corrections.

The imaginary part of the self-energy Fig. \ref{f.2b} is given by
the following expression
\begin{eqnarray}
&&\frac{\hbar}{\tau_T(\omega)} =\label{ret} {\rm Im}
\Sigma_{T}(\omega) = (\alpha j_2)^{2}\times \label{21} \\
&& \int\int d\varepsilon_1 d\varepsilon_2 {\rm
Im}K_{\sigma_l\sigma_2}^{(R)}(\varepsilon_1-\varepsilon_2) {\rm
Im}D(\omega-\varepsilon_1) {\rm Im}
G_S^{(R)}(\varepsilon_2)\nonumber
\\
&&\left(\tanh\frac{\varepsilon_2}{2T}+\coth\frac{\varepsilon_1-\varepsilon_2}{2T}\right)
\left(\tanh\frac{\varepsilon_1}{2T}+\coth\frac{\omega-\varepsilon_1}{2T}\right).
\nonumber
\end{eqnarray}
We assume that the decrement $\gamma$ in the propagator $D(t)$
(\ref{demp}) is small in comparison with the energies
$\Delta_{ST}$ and $\Delta_{ET}$ since we remain in the weakly
non-adiabatic regime. Here $ K^{R}(\omega) $ is the Fourier
transform of the spin susceptibility of the electron gas in the
leads
$$
K^{R}(t-t')=\sum_{k_1k_2k_3k_4}\langle
c^\dag_{k_1\sigma_1}(t)c_{k_2\sigma_2}(t)
c^\dag_{k_3\sigma_2}(t')c_{k_4\sigma_1}(t')\rangle_R
$$
which may be expressed via convolution of two electron
propagators,
$$
K^{R}(t-t')=\sum_{kq}g_{k\sigma_1}(t-t')g_{k+q,\sigma_2}(t'-t).
$$
The actual frequency interval is very narrow in comparison with
the electron Fermi energy, $\omega \ll \varepsilon_F$). At low
frequencies the Fourier transform of the electron-hole
polarization loop $K^{R}(\omega)$ behaves as
$$
{\rm Im}\; K^{R}(\omega)\sim \rho_0\frac{\omega}{\varepsilon_F} ~.
$$
Using this relation, one may estimate the dephasing rate:
\begin{eqnarray}
\gamma_2=\frac{\hbar}{2\tau_T(\omega \to 0)}\sim \left({\alpha
j_2}\right)^2 \left\{
\begin{array}{ll}
\displaystyle \Delta_{ST} e^{-\Delta_{ST}/T},~~ T < \Delta_{ST}
\nonumber \\
\\
\displaystyle \frac{T^2}{\Delta_{ST}},~~ \Delta_{ST} < T <
\Delta_{ET}
\end{array}\right.
\end{eqnarray}
We see that this contribution to the dephasing effect is frozen
out at low temperatures because the singlet states responsible for
dephasing is depopulated at $T\ll \Delta_{TS}$.

Now the self-energy corrections to the spin-fermion propagator
$G_{T\mu}$  represented by the diagrams Fig. \ref{f.0} and Fig.
\ref{f.2b} may be used in the scaling equations (\ref{scale}). As
a result of fluctuation corrections, $G_{T\mu}$ acquires the form
\begin{equation}\label{gew}
G_{T\mu}(\omega)=(\omega+i\gamma)^{-1},
\end{equation}
where $\gamma=\gamma_1$ at $T\to 0$ and $\gamma=\gamma_2$ at $T\to
\Delta_{ST}$. These propagators should be inserted in the diagrams
arising in the poor man's scaling procedure \cite{anders} (the
first of these diagrams is shown in Fig. \ref{f.2a}). Then we
immediately conclude that this imaginary part transforms into the
infrared cutoff of Kondo singularity. This means that there is no
significant effect of fluctuations on Kondo tunneling at high
$T\sim \Delta_{ST}$ provided $\gamma_2 \ll T_K$, where
$\gamma_2\sim T^2/\Delta_{ST}$. However both dephasing and
decoherence prevent the achievement of the unitarity limit at $T \to 0$.

Our last task is to clarify the contribution  of vertex
corrections presented in Figs. \ref{f.2a}b, \ref{ap.1} and
\ref{ap.2} to the scaling equations (\ref{scale}). It is clear
that the real parts of these corrections only slightly renormalize
the coupling parameters $j_{1,2}$ and do not influence the scaling
trajectories. As to the imaginary part $\Gamma^{''}_{TT}$, it
determines the longitudinal $1/T_1$ and transverse $1/T_2$
relaxation rates by means of the correlation function,
\begin{equation}\label{super_s}
\chi_{t}(\omega)=\langle {\bf S},{\bf S}\rangle_R
\end{equation}
which can be interpreted as line-shape. While all diagrams Fig.6
a-i contribute both to $1/T_1$ and $1/T_2$ relaxation rates, the
diagram Fig.7.a corresponds only to scattering processes without
spin flip, resulting in slightly different temperature behavior of
longitudinal and transverse relaxation rates. We note, however,
that the difference between $1/T_1$ and $1/T_2$ appears only
beyond the leading logarithmic approximation.

The imaginary part $\Gamma^{''}_{ST}$ is associated with spin
relaxation processes determined by the specific kind of dynamical
spin susceptibility
\begin{equation}\label{super}
\chi_{st}(\omega)=\langle {\bf P},{\bf P}\rangle_R~~,
\end{equation}
which describes the response of a magnetic system with $SO(n)$
symmetry not to external magnetic field but to the perturbations
intermixing triplet and singlet components of the spin manifold
[including gauge fluctuations (\ref{phasis1})]. This correlator,
given by  the irreducible S/T loop (see Fig.7.b-d) leads to the
appearance of a new $1/T_3$ relaxation rate associated with the
inverse time of transitions between the singlet and the 3-fold
degenerate triplet state.

The diagrams (a,b,d,e,g,h) presented in the first two columns of
Fig. \ref{ap.1} are in fact taken into account in the RG equations
together with the diagram (Fig. \ref{f.2a}b). Each of the three
diagrams (c,f,g) in the last column of Fig. \ref{ap.1} contains
two singlet propagators $G_S(\epsilon)$, so one should consider
them together with the non-parquet diagram (Fig. \ref{ap.2}a).
This vertex correction to $J_0^{TT}$ together with non-logarithmic
corrections (Fig. \ref{ap.2}b-d) due to fluctuation induced $(T\to
S \to T...)$ transitions accompany the adiabatic $TT$- and
$ST$-exchanges and introduce the longitudinal and transversal spin
relaxation times in the RG procedure (cf. Refs.
\onlinecite{KNG,Paas}).

The fluctuation induced vertex correction (Fig. \ref{ap.2}a) may
be estimated with the help of the Ward identity for $G_S(\omega\to
0)$ written in the form
\begin{equation}\label{ward}
\left. \frac{\partial \Sigma_{TT}}{\partial
\Delta_{TS}}\right|_{\omega \to 0}= \Gamma_{TT}(0),
\end{equation}
where $\Sigma_{TT}$ is the self energy shown in Fig. \ref{f.2b}
and $\Gamma_{TT}$ is the triplet vertex (Fig. \ref{ap.2}.(a)). The
Ward identity in this context corresponds to spin conservation in
the process of quasi-elastic scattering. We conclude from these
estimates that the non-adiabatic corrections $\hbar/(2\tau_{TT})=
Im \Gamma_{TT}$ presented by the diagram Fig. \ref{ap.2}.(a) are
exponentially weak at low $T\ll \Delta_{ST}$ similarly to the
dephasing rate $\gamma_2$.

Let us now consider the fluctuation corrections to the vertex
$J^{ST}$ represented in Fig. \ref{ap.2}.b-d. There is no Ward
identity for this inelastic process, which does not conserve spin
projection, so we calculate the vertex $\Gamma_{ST}$
straightforwardly with the help of Eq.(\ref{B6}). The imaginary
part of the diagram Fig. \ref{ap.2}b gives the following
correction to $\Gamma_{ST}(\omega,0)$  at $\omega \approx
\Delta_{ST}$ (see Appendix B):

\begin{eqnarray}
{\rm Im} [\rho_0\Gamma_{ST}^{(b)}]=\left(\alpha^2 j_2^3\right)
\left\{
\begin{array}{ll}
\displaystyle  T/\Delta_{ST},~~ \Delta_{ST} < T<\Delta_{ET}\\
\\
\displaystyle   \omega/\Delta_{ST},~~ T < \Delta_{ST}
\end{array}\right.
\end{eqnarray}
The estimates of the next diagrams Fig. \ref{ap.2}c,d  give a similar
result. Such behavior of non-diagonal vertex corrections is
predetermined by the threshold character of T/S excitations at finite
frequency. \cite{kis03}

These vertex corrections should be inserted in the scaling
equations (\ref{scale}). The imaginary part of the exchange vertex
introduces an additional cutoff in the scaling procedure.
\cite{Paas} Taking into account the fact that the contribution of
the singlet state to the flow equations controlled by the vertex
$j_2$ is frozen for $D < \Delta_{ST}$, one immediately finds that
the relaxation processes practically do not influence the cutoff
of $j_2$ because
\begin{equation}
\hbar /(2\tau_{ST})={\rm} {\rm Im} \Gamma_{ST}(\omega=\Delta_{ST})
\ll \Delta_{ST}.
\end{equation}

The real part
\begin{equation}
{\rm Re} [\rho_0\Gamma_{ST}]\sim \left(\alpha^2
j_2^3\right)(T/\Delta_{ST})\ln(D/T),
\end{equation}
just slightly disturbs the flow trajectories. Hence, it may also be
neglected in perturbative estimates.

As in the case of non-equilibrium Kondo effect\cite{Paas}, the
spin-relaxation processes are controlled by the imaginary part of
the susceptibility $K^R(\omega)$ of the electron liquid in
the leads. However, in our case the susceptibility is moderated by
the gauge fluctuations (broken line insertions in the
electron-hole or singlet-triplet loops in Fig. \ref{ap.2}(a-d)).
As a result, the relaxation corrections do not affect Kondo
tunneling at least in the weakly non-adiabatic regime.

\section{Conclusion}

A double quantum dot with a weak trembling potential applied to
the right dot (Fig. \ref{f.00}a) turned out to be an excellent
model system in which all facets of decoherence phenomenon are
exposed as observable effects. From a theoretical point of view,
this system is especially attractive because decoherence,
dephasing and relaxation are induced by the same gauge
fluctuations, which develop in the constrained Hilbert space of
the spin manifold $\{T,S,E\}$ (Fig. \ref{f.00}b) coupled to a Fermi
bath of conduction electrons. All these processes may be discussed
in a general context of the theory of decoherence in quantum
systems in contact with a thermal bath.

The {\it decoherence effect} characterized by the time $\tau_d$
(\ref{tau}) is related to the structure of the ground state of
TDQD in contact with the Fermi bath. It may be interpreted in
terms of the Superselection Rule introduced by Wick, Wightman and
Wigner for the description of baryonic charge (see Ref.
\onlinecite{Wight} for a recent review). Indeed, there is no
symmetry ban for superposition of two singlet states $|S\rangle$
and $|E\rangle$, but this superposition arises only as a result of
the coupling of the TDQD with the Fermi bath. The covering group,
which describes the symmetry of the manifold $\{T,S,E\}$ is
$SO(5)$, and the dynamical superposition is controlled by $U(1)$
gauge fluctuations (\ref{4.1}) under the Casimir constraint
(\ref{const2}). Due to these fluctuations, the coherent Kondo-type
ground state of the system 'TDQD + Fermi bath' cannot be reached.

The {\it dephasing effect} characterized by the time $\tau_{T}$
stems from the phase averaging in thermodynamical ensemble at
finite temperature. Dephasing $\theta_{ST}, \theta_{ET}$
(\ref{phasis}) emerge in a process of exchange scattering induced
by the random trembling potential. The scattering probabilities
are added incoherently, so that the spin-fermion self energy
acquires an imaginary part (Fig. \ref{f.2b}). Such processes are
generally classified as decoherence induced by dressing of bare
states.\cite{Zeh}

The {\it relaxation effects} characterized by the times
$\tau_{TT}$ and $\tau_{ST}$ remind those known in the conventional
theory of spin relaxation. Although we describe them in terms of
triplet-triplet and triplet-singlet transitions, one may
reinterpret the same processes in terms of longitudinal,
transversal and S/T relaxation rates $1/T_1$, $1/T_2$ and $1/T_3$
because both $TT$ and $TS$ processes contain spin conserving and
spin reversal components [see Eqs. (\ref{fff})]. However, since
all these processes are controlled by the small coupling constant
$\left({\alpha j_2}\right)^2$ induced by the trembling potential,
the relaxation contribution to dephasing processes is ineffective.

To conclude, we have shown in this paper that charge fluctuations
can be transformed into spin fluctuations, which result in
decoherence and dephasing of the Kondo effect in the double
quantum dot system. All these phenomena arise due to intrinsic
dynamical $SO(5)$ symmetry of the spin multiplet. Although we
considered only a weakly non-adiabatic regime where decoherence
effects are small by definition, the results are instructive,
because one may strictly discriminate between pure decoherence of
the ground state and the finite temperature dephasing and
relaxation effects in a situation, where all these effects are due
to gauge fluctuations within a spin multiplet. Strongly
non-adiabatic response to trembling gate potential demands special
consideration.
\section*{ACKNOWLEDGMENTS}

This work is partially supported by the SFB-410 and ISF grants. MK
acknowledges support through the Heisenberg program of the DFG.
Part of this work was done during the stay of MK, KK and JR
in the Max Planck Institute of Complex Systems, Dresden. KK benefited
from a visiting Professor position at the Universit\'e Louis Pasteur.
MK is grateful to ANL for the hospitality during his visit. Research
in Argonne was supported by U.S. DOE, Office of Science, under
Contract No. W-31-109-ENG-39.

\appendix
\section{Lie algebra for asymmetric DQD}

If the excitonic state is included in the set of the energy
levels, then the transitions between different states are
described by the $o_5$ algebra. In addition to standard $S=1$
operators
\begin{eqnarray}
S^+ & = & \sqrt{2}\left(X^{10}+X^{0,-1}\right),~
S^-  =  \sqrt{2}\left(X^{01}+X^{-1,0}\right),\nonumber \\
S_z & = & X^{11}-X^{-1,-1}, \label{A.1}
\end{eqnarray}
one should introduce two more vectors. The vector ${\bf P}$ with
the spherical components
\begin{eqnarray}
P^+ & = & \sqrt{2}\left(X^{1S}-X^{S,-1}\right),\;
P^-  = \sqrt{2}\left(X^{S1}-X^{-1,S}\right),\nonumber \\
P_z & = & -\left(X^{0S}+X^{S0}\right). \label{A.2}
\end{eqnarray}
defines transitions between the singlet $|S\rangle$ and the
components $|T\mu\rangle$ of spin triplet. Similarly, the vector
${\bf M}$ with components
\begin{eqnarray}
M^+ & = & \sqrt{2}\left(X^{1E}-X^{E,-1}\right),\;
M^-  = \sqrt{2}\left(X^{E1}-X^{-1,E}\right),\nonumber \\
M_z & = & -\left(X^{0E}+X^{E0}\right). \label{A.3}
\end{eqnarray}
determines transitions between the left exciton and the triplet.
The $o_5$ algebra is completed by the operator $A$
\begin{equation}
A = -i(X^{ES}-X^{SE}) .
\end{equation}
The Lie algebra is defined by the following commutation relations
\begin{equation}
[S_j,S_k]  = ie_{jkl}S_l,\;\;[P_j,P_k]=ie_{jkl}S_l,\;\;
[P_j,S_k]=ie_{jkl}P_l, \nonumber
\end{equation}
\begin{equation}
[M_j,M_k]  = ie_{jkl}S_l,\;\; [M_j,S_k]=ie_{jkl}M_l,\;\;
[P_j,M_k]=iA\delta_{jk}, \nonumber
\end{equation}
\begin{equation}
[P_j,A]  = iM_j,\;\; [A,M_j]=iP_j,\;\; [S_j,A]=0. \label{3.9e}
\end{equation}
($j,k,l$ are Cartesian coordinates). Besides,
\begin{equation}
{\bf S\cdot P} = 0,\;\;\; {\bf S\cdot M} = 0, \label{3.11}
\end{equation}
and the Casimir operator

\begin{equation}\label{A.4}
{\bf S}^2+ {\bf P}^2 + {\bf M}^2 + A^2=4.
\end{equation}

It is important to remind once more that the Hamiltonian $H_{dot}$
does not commute with vectors $\bf P$ and $\bf M$ because
\begin{eqnarray}
&&[P_z, {\bf P}^2] =  -[P_z, {\bf S}^2]= 2(X^{S0}-X^{0 S}),\nonumber\\
&&[P^+, {\bf P}^2] =  -[P^+, {\bf S}^2]=  4(X^{1S}+X^{S1})
\label{3.12}
\end{eqnarray}

The fermionic representation of $SO(5)$ group \cite{Nova,kisrew}
is characterized by 5-vector ${\bf q}^T=(f_{-1}^\dagger
f_0^\dagger, f_1^\dagger, f_s^\dagger, f_e^\dagger)$
\begin{eqnarray}
S^+ &=&  \sqrt{2}(f_0^\dagger f_{-1}+f^\dagger_{1}f_0),\;\;\;\;
S^z =
f^\dagger_{1}f_{1}-f^\dagger_{-1}f_{-1},\nonumber \\
P^+  &=& \sqrt{2}(f^\dagger_{1} f_s -  f_s^\dagger f_{-1}),
\;\;\;\; P^z = -( f_0^\dagger f_s + f_s^\dagger f_0),\nonumber\\
M^+  &=& \sqrt{2}(f^\dagger_{1} f_e -  f_e^\dagger f_{-1}),
\;\;\;\;  M^z = -( f_0^\dagger f_e + f_e^\dagger f_0).
\label{fff}\nonumber \\
A &=& i(f_e^\dagger f_s - f_s^\dagger f_e)
\end{eqnarray}
where $f^\dagger_{1}$, $f^\dagger_{\bar{1}}$ denote a creation
operator of the fermion with spin ``up'' and ``down'' respectively
whereas $f_0,f_s$,  stands for spinless fermions. Then the Casimir
operator (\ref{A.4}) transforms into the constraint
\begin{equation}
n_1+n_0+n_{-1}+n_s+n_r=1 \label{const2}
\end{equation}
\begin{widetext}
\section{Vertex corrections and Ward identities}
We start with the classification of the vertex corrections
containing one fluctuator line. The leading parquet diagrams are
plotted on Fig.\ref{ap.1}.a-i. These diagrams belong to three
different topological classes drawn on lines 1-3 of
Fig.\ref{ap.1}. Following the standard Feynman codex we write the
expressions for diagrams Fig.\ref{ap.1}.a-c

\begin{figure}[h]
\includegraphics[width=14cm,angle=0]{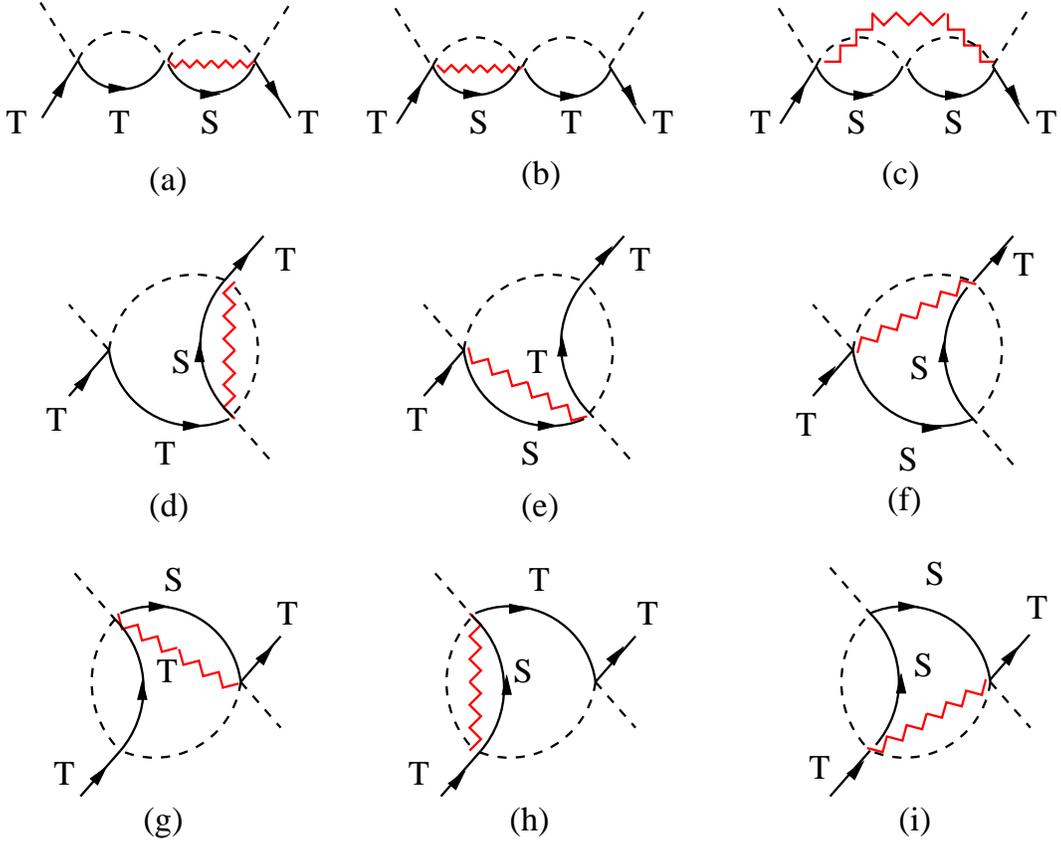}
\caption{First leading parquet corrections to TT vertex.  The
mapping of TT vertices on TS vertices is discussed in the text.
Similarly to Fig.5 the absence of direction in electron propagator
lines assumes that both directions (clockwise and anti-clockwise)
are possible.}\label{ap.1}
\end{figure}
$$
\Gamma_{TT} (\omega,\epsilon)\sim(\alpha j_2)^2 J_0^{\beta\beta'}
T^3\sum_{\sigma_1\sigma_2\mu_1\mu_2}R^{\sigma\sigma_1\sigma_2\sigma'}_{\mu\mu_1\mu_2\mu'}\sum_{\epsilon_1\epsilon_2\epsilon_3}\sum_{\bf
k_1k_2}
G_{\beta\mu_1}(\omega-\epsilon+\epsilon_1)G_{\beta'\mu_2}(\omega-\epsilon+\epsilon_2-\epsilon_3)g_{\sigma_1}({\bf
k_1},\epsilon_1)g_{\sigma_2}({\bf k_2},\epsilon_2)D(\epsilon_3)
$$
($\beta=T,S$). This expression corresponds to Fig.\ref{ap.1}a,
other expressions for diagrams b-c belong to the same topological
structure and differ by indices only. The summation is taken with
respect to fermionic Matsubara frequencies
$\epsilon_1,\epsilon_2=2i \pi T(n+1/2)$ and bosonic frequency
$\epsilon_3=2i \pi T n$.  Here the bare vertex $J^{SS}$, which
enters the diagrams (c,f,i), corresponds to the term
$J^{SS}\sum_{kk'}\sum_{\sigma}X^{SS}c^\dag_{k\sigma}
c^{}_{k'\sigma}$ in the potential scattering term $H_{ps}^{(0)}$
omitted in the SW Hamiltonian (\ref{SW1}). The tensor
$R^{\sigma\sigma_1\sigma_2\sigma'}_{\mu\mu_1\mu_2\mu'}$ stands for
kinematic factors containing ${\bf S}$ and ${\bf P}$ operators in
a scalar product with Pauli matrices $\hat \sigma$ and is given
by, for example,
\begin{equation}
R^{\sigma\sigma_1\sigma_2\sigma'}_{\mu \mu_1 s \mu'}= ({\bf
S}_{\mu_1\mu}\cdot {\mbox{\boldmath $\sigma$}}_{\sigma_1\sigma})
({\bf P}_{s \mu_1}\cdot {\mbox{\boldmath
$\sigma$}}_{\sigma_2\sigma_1})({ \bf P}_{\mu' s}\cdot
{\mbox{\boldmath $\sigma$}}_{\sigma'\sigma_2}),
\end{equation}
\begin{equation}
R^{\sigma\sigma_1\sigma_2\sigma'}_{\mu \mu_1 s \mu'}= ({\bf
S}_{\mu_1\mu}\cdot {\mbox{\boldmath $\sigma$}}_{\sigma_1\sigma})
({\bf P}_{s \mu_1}\cdot {\mbox{\boldmath
$\sigma$}}_{\sigma'\sigma_2})({\bf P}_{\mu' s}\cdot
{\mbox{\boldmath $\sigma$}}_{\sigma_1\sigma_2}),
\end{equation}
\begin{equation}
R^{\sigma\sigma_1\sigma_2\sigma'}_{\mu \mu_1 s \mu'}= ({\bf
S}_{\mu_1\mu}\cdot {\mbox{\boldmath $\sigma$}}_{\sigma_1\sigma_2})
({\bf P}_{s \mu_1}\cdot {\mbox{\boldmath
$\sigma$}}_{\sigma'\sigma_1})({ \bf P}_{\mu' s}\cdot
{\mbox{\boldmath $\sigma$}}_{\sigma_2\sigma})
\end{equation}
for Fig.6.a,d,g respectively. We assume also that the electron is
taken on the mass shell while $\epsilon$ is an energy transfer. The
energy of the triplet state is also assumed to be small
$\omega\ll\Delta_{ST}$.

For diagrams Fig.\ref{ap.1}.d-f one obtains
$$
\Gamma_{TT} (\omega,\epsilon)\sim(\alpha j_2)^2 J_0^{\beta\beta'}
T^3\sum_{\sigma_1\sigma_2\mu_1\mu_2}R^{\sigma\sigma_1\sigma_2\sigma'}_{\mu\mu_1\mu_2\mu'}\sum_{\epsilon_1\epsilon_2\epsilon_3}\sum_{\bf
k_1k_2}
G_{\beta\mu_1}(\omega-\epsilon+\epsilon_1)G_{\beta'\mu_2}(\omega-\epsilon+\epsilon_1-\epsilon_2-\epsilon_3)g_{\sigma_1}({\bf
k_1},\epsilon_1)g_{\sigma_2}({\bf k_2},\epsilon_2)D(\epsilon_3)
$$
and for Fig.\ref{ap.1}.g-i one gets
$$
\Gamma_{TT} (\omega,\epsilon)\sim(\alpha j_2)^2 J_0^{\beta\beta'}
T^3\sum_{\sigma_1\sigma_2\mu_1\mu_2}R^{\sigma\sigma_1\sigma_2\sigma'}_{\mu\mu_1\mu_2\mu'}\sum_{\epsilon_1\epsilon_2\epsilon_3}\sum_{\bf
k_1k_2}
G_{\beta\mu_1}(\omega-\epsilon+\epsilon_1-\epsilon_3)G_{\beta'\mu_2}(\omega-\epsilon+\epsilon_1-\epsilon_2)g_{\sigma_1}({\bf
k_1},\epsilon_1)g_{\sigma_2}({\bf k_2},\epsilon_2)D(\epsilon_3)
$$
Applying the procedure of analytical continuation explained in
Appendix C and taking the limit $\epsilon\to 0$ one gets the following
estimate for the real part of the diagrams
Fig.\ref{ap.1}(a,b,d,e,g,h)
\begin{equation}
{\rm Re}[\rho_0 \Gamma_{TT}(\omega,0)] \sim (\alpha j_2)^2
j_1\ln\left(\frac{D}{\max[T,\omega]}\right)\ln\left(\frac{D}{\Delta_{ST}}\right)
\end{equation}
while for diagrams Fig.\ref{ap.1}(c,f,i)
\begin{equation}
{\rm Re} [\rho_0\Gamma_{TT}(\omega,0)] \sim (\alpha j_2)^2
j_s\ln^2\left(\frac{D}{\Delta_{ST}}\right)
\end{equation}
These corrections are however parametrically smaller than the main
Kondo diagram Fig. \ref{f.2a}a:
\begin{equation} {\rm Re}
[\rho_0\Gamma^K_{TT}(\omega,0)] \sim (j_1)^3
\ln^2\left(\frac{D}{\max[T,\omega]}\right)
\end{equation}
under realistic conditions $\alpha<1$ and $T_K\ll \Delta_{ST}$.

The imaginary part of all diagrams a-i has a threshold form for
$\omega \gg T$
\begin{equation}
{\rm Im} \Gamma_{TT}(\omega,0) \sim (\alpha j_2)^2
J_0^{\beta\beta'}F_{\beta\beta'}(|\omega|-\Delta_{ST},
T)\theta(|\omega|-\Delta_{TS})
\end{equation}
where $F_{\beta\beta'}(|\omega|-\Delta_{ST}, T\ll\omega)\sim
(|\omega|-\Delta_{ST})^\nu$, and $\nu=1,2$ (see Ref.\cite{kis03}
for details of calculations). This estimation is done with
accuracy $O(\exp\left(-\Delta_{ST}/T\right))$. In contrast, the
imaginary part for Kondo vertices is given  in the limit
$\omega\gg T$
\begin{equation}
{\rm Im} [\rho_0\Gamma^K_{TT}(\omega,0)] \sim (j_1)^3 {\rm
sign}\;(\omega)
\end{equation}
The topological structure of singlet-triplet vertices is the same
as on Fig.\ref{ap.1}. The estimation for these diagrams gives the
following expressions
\begin{equation}
{\rm Re} [\rho_0\Gamma_{TS}(\omega,0)] \sim (j_2)^3
\ln\left(\frac{D}{\max[T,\omega]}\right)\ln\left(\frac{D}{\Delta_{ST}}\right)
\end{equation}
and
\begin{equation}
{\rm Im} [\rho_0\Gamma_{TS}(\omega,0)] \sim (\alpha^2 j_2^3
)(|\omega|-\Delta_{ST}) \theta(|\omega|-\Delta_{TS})
\end{equation}
Next to leading (non-parquet) diagrams are shown on
Fig.\ref{ap.2}.
\begin{figure}[h]
\includegraphics[width=12cm,angle=0]{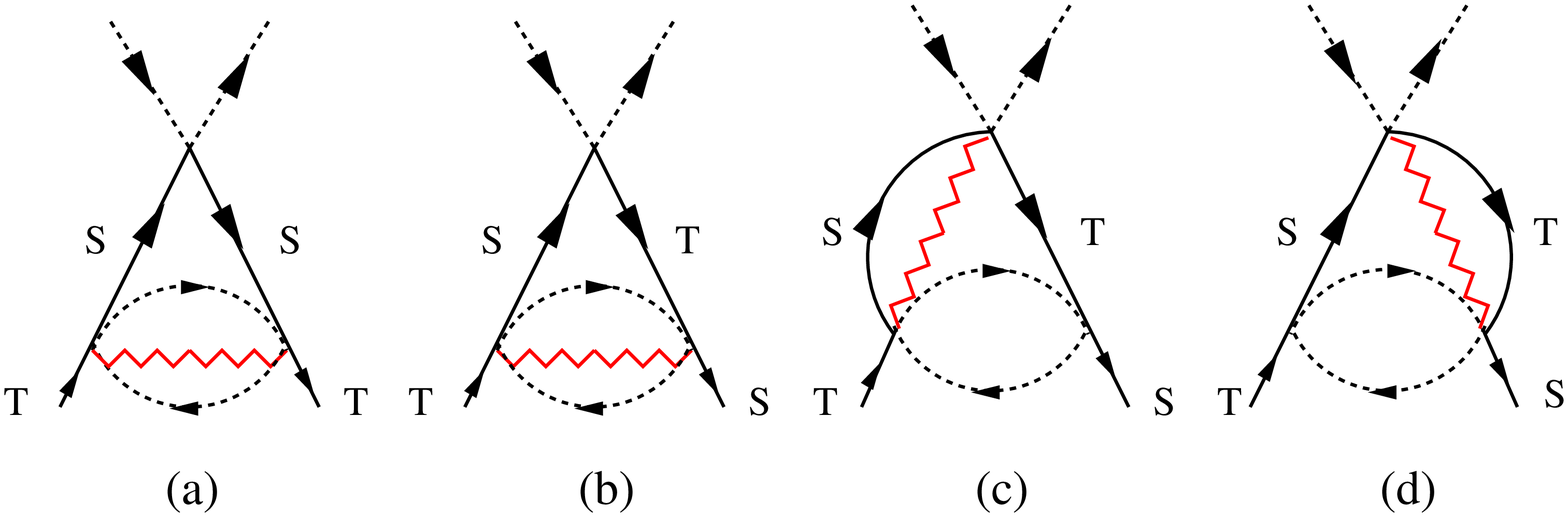}
\caption{Next to leading (non-parquet) corrections to the
vertices.}\label{ap.2}
\end{figure}

The analytical expression for diagram Fig.\ref{ap.2}.a (and
similarly for b) is given by
$$
\Gamma_{TT} (\omega,\epsilon)\sim(\alpha j_2)^2 J_0^{SS}
T^3\sum_{\sigma_1\sigma_2\mu_1\mu_2}R^{\sigma\sigma_1\sigma_2\sigma}_{\mu\mu}
\sum_{\epsilon_1\epsilon_2\epsilon_3}\sum_{\bf k_1k_2}
G_{S}(\epsilon+\epsilon_2)G_{S}(\epsilon_2)g_{\sigma_1}({\bf
k_1},\epsilon_1)g_{\sigma_2}({\bf
k_2},\epsilon_1+\epsilon_3-\omega)D(\epsilon_3)
$$
and for Fig.\ref{ap.2}.c (and similarly for d)
$$
\Gamma_{TS} (\omega,\epsilon)\sim(\alpha j_2)^2 J_0^{TS}
T^3\sum_{\sigma_1\sigma_2\mu_1\mu_2}R^{\sigma\sigma_1\sigma_2\sigma}_{\mu\mu_1\mu}
\sum_{\epsilon_1\epsilon_2\epsilon_3}\sum_{\bf k_1k_2}
G_{T\mu_1}(\epsilon+\epsilon_2)G_{S}(\epsilon_2)g_{\sigma_1}({\bf
k_1},\epsilon_1)g_{\sigma_2}({\bf
k_2},\epsilon_1+\epsilon_2+\epsilon_3-\omega)D(\epsilon_3)
$$
The easiest way to calculate diagram Fig.\ref{ap.2}.a is to use
the Ward Identity. The corresponding estimation is given in
Section IV. As one expected, the imaginary parts of these vertex
corrections have also threshold form ($|\omega| > \Delta_{ST}$).
\begin{eqnarray}
{\rm Im}[\rho_0\Gamma_{ST}^{(b)}]=\left(\alpha^2 j_2^3\right)
\left\{
\begin{array}{ll}
\displaystyle T/\Delta_{ST},~~ \Delta_{ST} < T<\Delta_{ET}
\nonumber \\
\\
\displaystyle \omega/\Delta_{ST},~~ T < \Delta_{ST}
\end{array}\right.
\end{eqnarray}
The damping $\hbar/\tau$ in the frequencies interval
$|\omega|<\Delta_{ST}$ is exponentially small ($\sim
\exp(-\Delta_{ST}/T)$) being proportional to the population of the
singlet state.

The real parts of TS vertices Fig.\ref{ap.2}.c-d are estimated as
\begin{equation}
{\rm Re} [\rho_0 \Gamma_{ST}(\omega,0)] \sim (\alpha^2 j_2^3)
\ln\left(\frac{D}{\Delta_{ST}}\right)
\end{equation}
The perturbative results including analysis of leading log and
sub-leading log diagrams summarized in this section allows to
incorporate noise corrections associated with the vertex fluctuator
into standard one- and  two- loops RG approach.
\section{Analytical continuation}
We start with the derivation of the general equation for the
vertex corrections $\Gamma_{\alpha\beta}(\omega,\epsilon)$ :
\begin{eqnarray}\label{B1}
\Gamma_{\alpha\beta}(\omega,\epsilon)= J^{\alpha\alpha'}
J^{\alpha'\beta'}J^{\beta'\beta} \int\frac{dz}{2\pi}\left[ {\rm Im
} G^R_{\alpha'}(z) G_{\beta'}^R(z+\epsilon)\Pi^R(\omega-z)
\tanh\left(\frac{z}{2T}\right)+\right. \nonumber\\
G^A_{\alpha'}(z) {\rm Im }G_{\beta'}^R(z+\epsilon)\Pi^R(\omega-z)
\tanh\left(\frac{z+\epsilon}{2T}\right) + \left.G^R_{\alpha'}(z)
G_{\beta'}^R(z+\epsilon){\rm Im }\Pi^R(\omega-z)
\coth\left(\frac{\omega-z}{2T}\right)\right]
\end{eqnarray}
Here the arguments in the vertex are complex variables defined in
the upper half-plane,
\begin{equation}\label{B1a}
\Pi(z)=
\int\frac{d\epsilon}{2\pi}\coth\left(\frac{\epsilon}{2T}\right)\left[
K^R(z-\epsilon){\rm Im} D^R(\epsilon)+D^R(z-\epsilon) {\rm Im}
K^R(\epsilon)\right].
\end{equation}
Then the elastic diagonal vertex $\Gamma_{TT}$ reads
\begin{eqnarray}\label{B2}
\Gamma_{TT}(\omega,0)= \left(J^{ST}\right)^2 J^{SS}
\int\frac{dz}{2\pi}\left[ \left(G^R_{S}(z)\right)^2{\rm Im
}\Pi^R(\omega-z) \coth\left(\frac{\omega-z}{2T}\right)+ 2{\rm
Re}G^R_{S}(z) {\rm Im }G_{S}^R(z)\Pi^R(\omega-z)
\tanh\left(\frac{z}{2T}\right) \right]\nonumber
\end{eqnarray}
This result may also be obtained by means of the Ward identity
(\ref{ward}) for the self energy insert $\Sigma_{ST}(\omega)$ in
the Green function $G_{T\mu_1\mu_2}$, which enters the triplet
vertex Fig. \ref{f.2b}.
\begin{equation}\label{B3}
\Sigma_{ST}(z)= \left(J^{ST}\right)^2
\int\frac{d\epsilon}{2\pi}\left[ {\rm Im } L^R(\epsilon)
D^R(z-\epsilon)
\tanh\left(\frac{\epsilon}{2T}\right)+L^R(z-\epsilon) {\rm Im}
D^R(\epsilon) \coth\left(\frac{\epsilon}{2T}\right)\right],
\end{equation}
where
\begin{equation}\label{B3a}
L^R(z)=\int\frac{d\epsilon}{2\pi}\left[ {\rm Im } G^R(\epsilon)
K^R(z-\epsilon)
\tanh\left(\frac{\epsilon}{2T}\right)+G^R(z-\epsilon) {\rm Im}
K^R(\epsilon) \coth\left(\frac{\epsilon}{2T}\right)\right]
\end{equation}
The imaginary part of Eq. (\ref{B1}) gives the expression
(\ref{21}) for the dephasing rate

No Ward identity exists for the non-diagonal vertex
\begin{eqnarray}\label{B6}
\Gamma_{TS}(\omega,\epsilon)= \left( J^{ST}\right)^3
\int\frac{dz}{2\pi}\left[ {\rm Im } G^R_{S}(z)
G_{T}^R(z+\epsilon)\Pi^R(\omega-z)
\tanh\left(\frac{z}{2T}\right)+\right. \nonumber\\
G^A_{S}(z) {\rm Im }G_{T}^R(z+\epsilon)\Pi^R(\omega-z)
\tanh\left(\frac{z+\epsilon}{2T}\right) + \left.G^R_{S}(z)
G_{T}^R(z+\epsilon){\rm Im }\Pi^R(\omega-z)
\coth\left(\frac{\omega-z}{2T}\right)\right]
\end{eqnarray}
\end{widetext}

\end{document}